\documentclass[10pt,a4paper,english,nofootinbib]{revtex4-2}
\usepackage{lmodern}
\usepackage{lmodern}

\usepackage[T1]{fontenc}
\usepackage[latin9]{inputenc}
\usepackage{babel}
\usepackage{calc}
\usepackage{amsmath}
\usepackage{amssymb}
\usepackage{graphicx}
\usepackage{esint}
\usepackage[unicode=true,pdfusetitle,
 bookmarks=true,bookmarksnumbered=false,bookmarksopen=false,
 breaklinks=false,pdfborder={0 0 1},backref=false,colorlinks=false]
 {hyperref}
\usepackage{xcolor}
\makeatletter

\@ifundefined{textcolor}{}
{%
 \definecolor{BLACK}{gray}{0}
 \definecolor{WHITE}{gray}{1}
 \definecolor{RED}{rgb}{1,0,0}
 \definecolor{GREEN}{rgb}{0,1,0}
 \definecolor{BLUE}{rgb}{0,0,1}
 \definecolor{CYAN}{cmyk}{1,0,0,0}
 \definecolor{MAGENTA}{cmyk}{0,1,0,0}
 \definecolor{YELLOW}{cmyk}{0,0,1,0}
}

\usepackage{babel}

\usepackage{graphicx}
\def\b{\begin{equation}}
\def\e{\end{equation}}

\@ifundefined{textcolor}{}{%
 \definecolor{BLACK}{gray}{0}
 \definecolor{WHITE}{gray}{1}
 \definecolor{RED}{rgb}{1,0,0}
 \definecolor{GREEN}{rgb}{0,1,0}
 \definecolor{BLUE}{rgb}{0,0,1}
 \definecolor{CYAN}{cmyk}{1,0,0,0}
 \definecolor{MAGENTA}{cmyk}{0,1,0,0}
 \definecolor{YELLOW}{cmyk}{0,0,1,0}
 }

\usepackage{latexsym}\usepackage{bm}

\makeatother

\begin{document}
\title{Hawking Temperature as the Total Gauss-Bonnet Invariant of the Region
Outside a Black Hole }
\author{Emel Altas}
\email{emelaltas@kmu.edu.tr}

\affiliation{Department of Physics,\\
 Karamanoglu Mehmetbey University, 70100, Karaman, Turkey}
\author{Bayram Tekin}
\email{btekin@metu.edu.tr}

\affiliation{Department of Physics,\\
 Middle East Technical University, 06800, Ankara, Turkey}
\date{\today}
\begin{abstract}
We provide two novel ways to compute the surface gravity ($\kappa$) and the Hawking temperature $(T_{H})$ of a stationary black hole: in the first method $T_{H}$ is given
as the three-volume integral of the Gauss-Bonnet invariant (or the
Kretschmann scalar for Ricci-flat metrics) in the total region outside
the event horizon; in the second method it is given as the surface
integral of the Riemann tensor contracted with the covariant derivative
of a Killing vector on the event horizon. To arrive at these new formulas
for the black hole temperature (and the related surface gravity), we first
construct a new differential geometric identity using the Bianchi
identity and an antisymmetric rank-$2$ tensor, valid for spacetimes with at least one Killing vector field. The Gauss-Bonnet tensor
and the Gauss-Bonnet scalar play a particular role in this geometric identity. We calculate
the surface gravity and the Hawking temperature of the Kerr and the extremal Reissner-Nordstr\"om
holes as examples. 
\end{abstract}
\maketitle

\section{Introduction}

Black hole physics, from the vantage point of both observations and
theory, is in a remarkable state of development. Rotating Kerr metric
\cite{Kerr:1963ud}, as the vacuum solution of Einstein field equations,
describe all the observed properties of these black holes with just
two parameters: the mass of the black hole $m$ and the rotation parameter
$a$ which is related to the angular momentum of the black hole as
$a=J/m$ ({See \cite{Frolov} for a detailed exposition.})
These two parameters arise as integration constants in the
solution of the partial differential equations; but they can be represented
as geometric invariants through the usual ADM \cite{adm} construction
which expresses the mass and angular momentum as integrals of the
first derivatives of the metric tensor at spatial infinity \cite{Eric0}. 
Besides these parameters, the black hole is expected to have thermal properties: for example,
an equilibrium black hole obeys the four laws of black hole physics
\cite{bardeen, bekenstein, israel}.  {{An explicitly gauge invariant derivation of the black hole laws of black hole thermodynamics was given recently in \cite{Shahin}}

Thermodynamics of black holes is currently only understood at a semi-classical level \cite{hawking}; and hence a proper microscopic understanding of this issue is important for quantum gravity.  In black hole thermodynamics, the notion of surface gravity, associated to a Killing (or event) horizon, plays a major role as it is directly related to the zeroth law and the uniform temperature assigned to a black hole. Surface gravity is usually defined as 
the nonaffinity coefficient ($\kappa$) in the null Killing vector field  $\zeta^{\mu}$ given as:
\begin{equation}
\zeta^{\mu}\nabla_{\mu}\zeta^{\nu}=-\kappa\zeta^{\nu},\label{surfacegravity}
\end{equation}
which is to be computed on the event horizon. {{We must keep in mind the well-known ambiguity in the definition of surface gravity here: if the integral curves of the null Killing vector $\zeta^\mu$ are restricted to be affinely parameterized, then $\nabla_\zeta\zeta=0$ and 
$\kappa$ disappears. So affine parameterization should not be imposed. Furthermore, a constant scaling of $\zeta^\mu \rightarrow a\zeta^\mu $, also scales $\kappa \rightarrow a \kappa$. So one must fix the normalization $\zeta^\mu$ away from the horizon where it is not null, an issue to which we shall come back below. For two wonderful expositions of this topic, see \cite{Wald, Eric}. As demonstrated in these works in a pedagogical manner, one can show that the surface gravity as defined above (\ref{surfacegravity}) is constant on the horizon which, even at a cursory level,  suggests a direct connection of the event horizon (or the black hole) with an object with constant temperature that is in equilibrium with its surrounding. }}

Here we provide a completely unexpected formulation of surface gravity which matches the usual formulation (\ref{surfacegravity}) for stationary black holes. {{Our definition is valid for generic spacetime dimensions larger than 3, and for generic gravity theories.}} But, in particular, for four dimensional vacuum black holes, we show that the surface gravity is proportional to the volume integral of the Kretschmann scalar outside the black hole region. It is well-known that the Kretschmann scalar diverges for a black hole in some region inside the event horizon and this scalar has been used to  detect the singularity of black hole spacetimes {{See \cite{Lake} for use of the scalar curvature invariants on the detection of other invariants of black holes. } Here we have shown another use of  Kretschmann curvature invariant: its integral over the spatial section of the spacetime outside the black hole yields the surface gravity and hence the associated Hawking temperature \cite{hawking} given in geometric units as
\begin{equation}
T_{H}=\frac{\kappa}{2\pi}.\label{Hawkingtemperature}
\end{equation}
 In what follows, we will show that $\kappa$ can be expressed
as the total three-volume integral of the Kretschmann scalar, where
the integration domain is outside the black hole region, that is
from the event horizon to spatial infinity of the spacetime. Equivalently,
it can also be expressed as a surface integral (see Fig. 1 (\ref{fig1})) on the
cross section of the event horizon with an integral that involves
the Riemann tensor and the covariant derivative of the timelike or
any other Killing vector. Our formulation is geometric in the sense that it is valid for any gravity theory, for any $n\ge 4$ dimensions. The contents of a theory enter only after the geometric identity (\ref{mainidentity}).

To derive the new formulas for surface gravity and the black hole temperature , let us start
with the construction of a new geometric identity.
\begin{figure}
~~\includegraphics[scale=0.7]{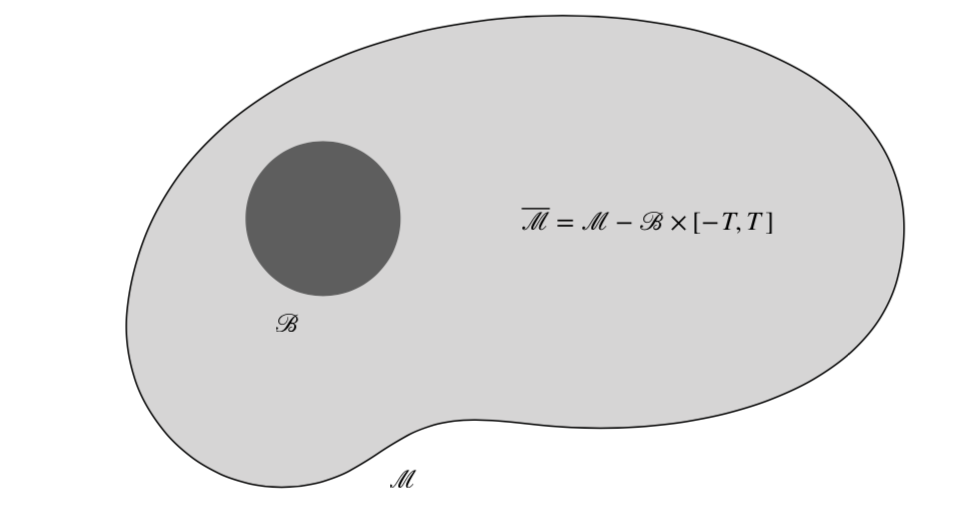}
\caption{$\mathcal{M}$ denotes the four {{(or generically $n>3$)}} dimensional spacetime, $\mathcal{B}$
represents the three {{(or generically $n-1$)}} dimensional ball for which the boundary is the
cross section of the event horizon. Also, $\bar{\mathcal{M}}=\mathcal{M}-\mathcal{B}\times\left[-T,T\right]$
denotes the region of the spacetime between the event horizon and
the boundary of the black hole at infinity. {{To not deal with a trivial divergence over the time integral for
stationary spacetimes, we have taken the time dimension to run over
the interval $\left[-T,T\right]$}}. The boundary of $\mathcal{\bar{M}}$,
$\partial\mathcal{\bar{M}}$, consists of the event horizon as a
$3$ (generically $(n-1)$) dimensional degenerate hypersurface and the boundary at infinity.}
\label{fig1}
\end{figure}

\section{Construction of the geometric identity }

In \cite{Emel_PRD_uzun, Altas}, we introduced the following
$\mathcal{P}$-tensor 
\begin{equation}
\text{\ensuremath{{\cal {P}}}}^{\nu}\thinspace_{\mu\beta\sigma}:=R^{\nu}\thinspace_{\mu\beta\sigma}+\delta_{\sigma}^{\nu}\text{\ensuremath{{\cal {G}}}}_{\beta\mu}-\delta_{\beta}^{\nu}\text{\ensuremath{{\cal {G}}}}_{\sigma\mu}+\text{\ensuremath{{\cal {G}}}}_{\sigma}^{\nu}g_{\beta\mu}-\text{\ensuremath{{\cal {G}}}}_{\beta}^{\nu}g_{\sigma\mu}
+(\frac{R}{2}-\frac{\Lambda(n+1)}{n-1})(\delta_{\sigma}^{\nu}g_{\beta\mu}-\delta_{\beta}^{\nu}g_{\sigma\mu}).
\end{equation}
where ${\cal {G}}_{\beta}^{\nu}:=R_{\beta}^{\nu}-\frac{1}{2}R\delta_{\nu}^{\beta}+\Lambda\delta_{\nu}^{\beta}$.
The $\mathcal{P}$-tensor (which vanishes identically in three dimensions)
satisfies the symmetries of the Riemann tensor and its contraction
yields the Einstein tensor, $\text{\ensuremath{{\cal {P}}}}^{\nu}\thinspace_{\mu v\sigma}=(3-n)\text{\ensuremath{{\cal {G}}}}_{\mu\sigma}$. {{In fact, one of our motivations was to find a rank $(1,3)$ tensor whose contraction is not the Ricci tensor, but the Einstein tensor. This $\mathcal{P}$-tensor does the job.}}
Moreover, as defined above, this tensor vanishes identically for
maximally symmetric spacetimes; but when $\Lambda=0$, it vanishes
for flat spacetimes. Perhaps, the most important property of the $\mathcal{P}$-tensor
is that, unlike the Riemann tensor, it is divergence-free for all twice
differentiable metrics on a spacetime
\begin{equation}
\nabla_{\nu}\ensuremath{{\cal {P}}}^{\nu\mu}\thinspace_{\beta\sigma}=0.\label{divergenceofp}
\end{equation}
This fact yields rather remarkable consequences for the underlying
manifold. Let $\text{\ensuremath{{\cal {F}}}}^{\beta\sigma}$ be a
generic {\it  antisymmetric tensor}. Then, contracting (\ref{divergenceofp})
with $\text{\ensuremath{{\cal {F}}}}^{\beta\sigma}$ yields 
\begin{equation}
\nabla_{\nu}(\ensuremath{{\cal {P}}}^{\nu\mu}\thinspace_{\beta\sigma}\mathcal{F}^{\beta\sigma})=\ensuremath{{\cal {P}}}^{\nu\mu}\thinspace_{\beta\sigma}\nabla_{\nu}\mathcal{F}^{\beta\sigma}.\label{exactequation}
\end{equation}
Let $\chi^{\sigma}$ be a generic vector field on the manifold. Then,
we take a particular $\mathcal{F}^{\beta\sigma}$ such that $\chi^{\sigma}$
be its potential as 
\begin{equation}
\mathcal{F}^{\beta\sigma}=\frac{1}{2}\left(\nabla^{\beta}\chi^{\sigma}-\nabla^{\sigma}\chi^{\beta}\right),
\end{equation}
and decompose $\chi^{\sigma}$ as follows 
\begin{equation}
\chi^{\sigma}:=\xi^{\sigma}+\psi^{\sigma},
\end{equation}
where $\xi^{\sigma}$ is a Killing vector ({ i.e. $\nabla^{\beta}\xi^{\sigma}+\nabla^{\sigma}\xi^{\beta}=0$}) and $\psi^{\sigma}$ is
a generic vector. Then $\mathcal{F}^{\beta\sigma}$ becomes 
\begin{equation}
\mathcal{F}^{\beta\sigma}=\nabla^{\beta}\xi^{\sigma}+\frac{1}{2}\left(\nabla^{\beta}\psi^{\sigma}-\nabla^{\sigma}\psi^{\beta}\right).\label{fbetasigma}
\end{equation}
Using the Killing identity, $\nabla_{\nu}\nabla^{\beta}\xi^{\sigma}=R^{\sigma\beta}\thinspace_{\nu\lambda}\xi^{\lambda}$ \cite{Weinberg},
the right hand side of equation (\ref{exactequation}) can be written
as 
\begin{equation}
\ensuremath{{\cal {P}}}^{\nu\mu}\thinspace_{\beta\sigma}\nabla_{\nu}\mathcal{F}^{\beta\sigma}=\ensuremath{{\cal {P}}}^{\nu\mu}\thinspace_{\beta\sigma}R^{\sigma\beta}\thinspace_{\nu}\thinspace^{\lambda}\xi_{\lambda}+\nabla_{\nu}(\ensuremath{{\cal {P}}}^{\nu\mu}\thinspace_{\beta\sigma}\nabla^{\beta}\psi^{\sigma}).\label{eq1}
\end{equation}
We will now write the contraction of the Riemann and the $\mathcal{P}$-tensors,
$\ensuremath{{\cal {P}}}^{\nu\mu}\thinspace_{\beta\sigma}R^{\sigma\beta}\thinspace_{\nu\lambda}$,
in terms of the Gauss-Bonnet tensor \cite{Gurses,gursesprl} 
\begin{align}
\mathcal{H}_{\mu\nu} & :=2\Bigl(2RR_{\mu\nu}-2R_{\mu\alpha\nu\beta}R^{\alpha\beta}+R_{\mu\alpha\beta\sigma}R_{\nu}\thinspace^{\alpha\beta\sigma}\nonumber \\
 & -2R_{\mu\alpha}R_{\nu}^{\alpha}-\frac{1}{4}g_{\mu\nu}\chi_{GB}\Bigr),
\label{gb}
\end{align}
where the Gauss-Bonnet invariant, $\chi_{GB}$, reads as 
\begin{equation}
\chi_{GB}:=R_{\mu\alpha\beta\sigma}R{}^{\mu\alpha\beta\sigma}-4R_{\mu\nu}R^{\mu\nu}+R^{2}.
\end{equation}
Note that $\mathcal{H}_{\mu\nu}$ vanishes identically in four dimensions,
while $\chi_{GB}$ can be written as the divergence of a vector field,
albeit in a non-covariant way. ({Vanishing of $\mathcal{H}_{\mu\nu}$ in four dimensions is not obvious from the definition (\ref{gb}), but a detailed derivation was given in \cite{Gurses} and a concise one in \cite{gursesprl}.})
Then in generic $n$ spacetime dimensions,
one has an identity
\begin{equation}
\ensuremath{{\cal {P}}}^{\nu\mu}\thinspace_{\beta\sigma}R^{\sigma\beta}\thinspace_{\nu}\thinspace^{\lambda}=-\frac{1}{2}\mathcal{H}^{\mu\lambda}-\frac{1}{4}g^{\mu\lambda}\chi_{GB}+\frac{2\Lambda(n-3)}{(n-1)}R^{\mu\lambda}.\label{eq2}
\end{equation}
Inserting (\ref{fbetasigma}, \ref{eq2}) in equation (\ref{exactequation}),
one arrives at the desired geometric identity which is valid for any smooth metric

\noindent\fbox{\begin{minipage}[t]{1\columnwidth - 2\fboxsep - 2\fboxrule}%
\begin{align}\label{mainidentity}
 & \nabla_{\nu}(\ensuremath{{\cal {P}}}^{\nu\mu}\thinspace_{\beta\sigma}\nabla^{\beta}\xi^{\sigma})=\Bigl(\frac{2\Lambda(n-3)}{(n-1)}R^{\mu\lambda}-\frac{1}{2}\mathcal{H}^{\mu\lambda}-\frac{1}{4}g^{\mu\lambda}\chi_{GB}\Bigr)\xi_{\lambda}. 
\end{align}
\end{minipage}}\\
\\
Note that, at the end of the construction, the non-Killing part of
the $\chi^{\sigma}$ vector dropped in the last equation, and only
the Killing part survived. So this identity is valid only for spacetimes
that have at least one Killing symmetry, otherwise one does not have
this identity.

{{As the identity (\ref{mainidentity}) is a vector identity, its prone to another covariant derivative. Let us show that, without a constraint on the geometry beyond the assumption of the existence of a Killing symmetry, the covariant derivative of the identity vanishes automatically. This really is desired, otherwise the underlying geometry would be further constrained.  So we have
\begin{align}\label{mainidentity22}
 & \nabla_{\mu}\nabla_{\nu}(\ensuremath{{\cal {P}}}^{\nu\mu}\thinspace_{\beta\sigma}\nabla^{\beta}\xi^{\sigma})=\nabla_{\mu}\Bigl(\frac{2\Lambda(n-3)}{(n-1)}R^{\mu\lambda}-\frac{1}{2}\mathcal{H}^{\mu\lambda}-\frac{1}{4}g^{\mu\lambda}\chi_{GB}\Bigr)\xi_{\lambda}.
\end{align}
}
{{
Let us concentrate on the left-hand side which reads
\begin{align}
 \nabla_{\mu}\nabla_{\nu}(\ensuremath{{\cal {P}}}^{\nu\mu}\thinspace_{\beta\sigma}\nabla^{\beta}\xi^{\sigma}) &= \frac{1}{2}  [\nabla_{\mu},\nabla_{\nu}](\ensuremath{{\cal {P}}}^{\nu\mu}\thinspace_{\beta\sigma}\nabla^{\beta}\xi^{\sigma})\nonumber \\
&=R_{\mu \nu}\,^\nu\,_\lambda (\ensuremath{{\cal {P}}}^{\lambda\mu}\thinspace_{\beta\sigma}\nabla^{\beta}\xi^{\sigma})\nonumber \\
&+R_{\mu \nu}\,^\mu\,_\lambda (\ensuremath{{\cal {P}}}^{\nu\lambda}\thinspace_{\beta\sigma}\nabla^{\beta}\xi^{\sigma})\nonumber \\
&=-R_{\mu \lambda}(\ensuremath{{\cal {P}}}^{\lambda\mu}\thinspace_{\beta\sigma}\nabla^{\beta}\xi^{\sigma})\nonumber \\
&~~~+R_{\nu \lambda} (\ensuremath{{\cal {P}}}^{\nu\lambda}\thinspace_{\beta\sigma}\nabla^{\beta}\xi^{\sigma}).
\end{align}
In the last two lines, each term vanished identically since the Ricci tensor is symmetric while the term in the parenthesis is anti-symmetric. So the lest-hand side of (\ref{mainidentity22}) vanishes identically. Let us check the right-hand side of that equation. Since $\xi$ is a Killing vector $\nabla_\xi \xi=0$.  Since the geometry is invariant along the flow of this Killing vector, we have $\nabla_\xi R= \xi^\mu\nabla_\mu R=0$, which can also be easily shown. Similarly $\nabla_\xi  \chi_{GB}=0$. As the Gauss-Bonnet tensor $\mathcal{H}^{\mu\lambda}$ comes from the variation of a diffeomorphism invariant action, it satisfies covariant conservation
$\nabla_\mu \mathcal{H}^{\mu\lambda}=0$ So the right-hand side of (\ref{mainidentity22}) boils down to $\frac{2\Lambda(n-3)}{(n-1)}\xi_\lambda \nabla_{\mu}R^{\mu\lambda}$ which vanishes identically upon use of the Bianchi Identity $\nabla_{\mu}R^{\mu\lambda}= \frac{1}{2}\nabla^\lambda R$ plus the  identity $\xi^\mu\nabla_\mu R=0$ coming from the Killing vector identity. So to some up: covariant derivative of (\ref{mainidentity}) vanishes for all smooth geometries. And as we have just shown, since the right-hand and the left-hand vanishes independently, identically, this allows us to define two equivalent covariantly conserved currents :
\begin{align}
{\mathcal{J}}^\mu :=  \nabla_{\nu}(\ensuremath{{\cal {P}}}^{\nu\mu}\thinspace_{\beta\sigma}\nabla^{\beta}\xi^{\sigma}) 
\end{align}
and
\begin{align}
{\mathcal{J}}^\mu =\Bigl(\frac{2\Lambda(n-3)}{(n-1)}R^{\mu\lambda}-\frac{1}{2}\mathcal{H}^{\mu\lambda}-\frac{1}{4}g^{\mu\lambda}\chi_{GB}\Bigr)\xi_{\lambda}.
\end{align}
We shall use both of these two give two different expressions for the surface gravity and temperature of a stationary black hole.
}

Since, up to now, we have not assumed any field equations, in principle we
can consider any gravity theory, but {{to derive the consequences of  (\ref{mainidentity}) for our World,}} let us consider four dimensional
manifolds that satisfy the cosmological Einstein theory with matter.
Then, one has $\mathcal{G}_{\mu\nu}=\kappa_N T_{\mu\nu}$, {{with $\kappa_N = \frac{ 8 \pi G_n}{c^4}$}; and as stated
above {{the Gauss-Bonnet tensor}} {$\mathcal{H}_{\mu\nu} = 0$ {{in four dimensions}}, yielding 
\begin{align}
 & \nabla_{\nu}(\ensuremath{{\cal {P}}}^{\nu\mu}\thinspace_{\beta\sigma}\nabla^{\beta}\xi^{\sigma})=\Bigl(-\frac{1}{4}g^{\mu\lambda}R_{\rho\alpha\beta\sigma}R^{\rho\alpha\beta\sigma}\nonumber \\
 & ~+\kappa_N^{2}g^{\mu\lambda}\widetilde{T}_{\alpha\beta}\widetilde{T}^{\alpha\beta}-\frac{1}{4}g^{\mu\lambda}\kappa_N^{2}\widetilde{T}{}^{2}+\frac{2\Lambda}{3}\kappa\widetilde{T}^{\mu\lambda}\Bigr)\xi_{\lambda},\label{denklem}
\end{align}
{{where we have expressed the right-hand side of the identity in terms of the(modified)} energy momentum tensor: $\widetilde{T}_{\mu\nu}:=T_{\mu\nu}-\frac{1}{2}g_{\mu\nu}T+\frac{\Lambda}{\kappa_N}g_{\mu\nu}$. {{In particular, one of the main applications of this construction will be the astrophysically relevant Kerr black hole for which } $\Lambda=0$ and in a vacuum, $T_{\mu\nu}=0$, and hence (\ref{denklem})
reduces to

\begin{equation}
{\mathcal{J}}^\mu =\nabla_{\nu}(\ensuremath{{\cal {P}}}^{\nu\mu}\thinspace_{\beta\sigma}\nabla^{\beta}\xi^{\sigma})=-\frac{1}{4}\xi^{\mu}R_{\rho\alpha\beta\sigma}R^{\rho\alpha\beta\sigma}.\label{last}
\end{equation}
{{ So on the right-hand side} the Gauss-Bonnet invariant reduced to the Kretschmann scalar for Ricci
flat metrics  {{i.e. the metrics solving vacuum Einstein equation}}. 

{{As we have shown in the general case above, above we have 
$\nabla_\mu{\mathcal{J}}^\mu=0$, which yields a true conservation law  
$\partial_\mu ( \sqrt{-g}{\mathcal{J}}^\mu)=0$ which can be integrated over the spacetime $\int_{\mathcal{M}} d^4 x \partial_\mu ( \sqrt{-g}{\mathcal{J}}^\mu)=0$. } 

Let us consider the consequences of this expression
for black hole spacetimes. {{ For a detailed discussion of this type of construction, please see the third section in \cite{Tah}. To be concrete, let $\xi^\mu$ be a time-like Killing vector and let $\Sigma$ be a spatial hypersurface (which will be specified below) in the total spacetime $\mathcal{M}$ and let $n_mu$ be its (inward-pointing) unit time-like normal vector, and $\gamma_{ij}$ be the induced metric on $\Sigma$.
Then $\int_\Sigma d^3 y \sqrt{\gamma} n_{\mu} {\mathcal{J}}^\mu$ is independent of time and the  choice of the spatial hypersurface as per conservation. 
So then we have the following exact relation }

\begin{eqnarray}\label{integralequation0}
 &  & \intop_{\Sigma }d^{3}y\sqrt{\gamma}\thinspace n_{\mu}\nabla_{\nu}(R^{\nu\mu}\thinspace_{\beta\sigma}\nabla^{\beta}\xi^{\sigma})=-\frac{1}{4}\intop_{\Sigma}d^{3}y\sqrt{\gamma}\,n_\mu \xi^{\mu}R_{\rho\alpha\beta\sigma}R^{\rho\alpha\beta\sigma},
\end{eqnarray}
\vskip 0.3 cm\noindent
{{where in vacuum, for Ricci flat metrics, the $\ensuremath{{\cal {P}}}$ reduced to the Riemann tensor. We can use the Stokes' theorem on the left-hand side as follows 
\begin{eqnarray}\label{integraleq2}
 &  & \intop_{\Sigma }d^{3}y\sqrt{\gamma}\thinspace n_{\mu}\nabla_{\nu}(R^{\nu\mu}\thinspace_{\beta\sigma}\nabla^{\beta}\xi^{\sigma})=\intop_{\partial\Sigma }d^{2}z\sqrt{\gamma^{(\partial \Sigma)}}\thinspace n_{\mu}\sigma_{\nu}R^{\nu\mu}\thinspace_{\beta\sigma}\nabla^{\beta}\xi^{\sigma}
\end{eqnarray}
where $\partial \Sigma$ is the (spacelike) boundary of the spacelike surface $\Sigma$ while $\sigma_\nu$ is its spacelike outward unit normal vector and $\gamma^{(\partial \Sigma)}_{\mu \nu} := g_{\mu \nu} + n_\mu n_\nu- \sigma_\mu \sigma_\nu$ is the induced metric on it.  Introducing the antisymmetric binormal as 
\begin{equation}
\epsilon_{\mu \nu} :=\frac{1}{2} \left (  n_{\mu}\sigma_{\nu}-n_{\nu}\sigma_{\mu}\right),
\end{equation}
we can rewrite (\ref{integraleq2}) as
\begin{eqnarray}\label{integraleq3}
 &  & \intop_{\Sigma }d^{3}y\sqrt{\gamma}\thinspace n_{\mu}\nabla_{\nu}(R^{\nu\mu}\thinspace_{\beta\sigma}\nabla^{\beta}\xi^{\sigma})=\intop_{\partial\Sigma }d^{2}z\sqrt{\gamma^{(\partial \Sigma)}}\epsilon_{\mu \nu}R^{\nu\mu}\thinspace_{\beta\sigma}\nabla^{\beta}\xi^{\sigma}.
\end{eqnarray}
So then we have the main identity
\vskip 0.3 cm
\noindent\fbox{\begin{minipage}[t]{1\columnwidth - 2\fboxsep - 2\fboxrule}%
\begin{eqnarray}\label{integralequation}
 &  &\intop_{\partial\Sigma }d^{2}z\sqrt{\gamma^{(\partial \Sigma)}}\epsilon_{\mu \nu}R^{\nu\mu}\thinspace_{\beta\sigma}\nabla^{\beta}\xi^{\sigma}=-\frac{1}{4}\intop_{\Sigma}d^{3}y\sqrt{\gamma}\,n_\mu \xi^{\mu}R_{\rho\alpha\beta\sigma}R^{\rho\alpha\beta\sigma}.
\end{eqnarray}
\end{minipage}}
\vskip 0.3 cm\noindent
}
{{To proceed, let us now specify the hypersurface $\Sigma$.  To be mathematically somewhat rigorous, let us take the time interval to be compact $t\in\left[-T,T\right]$ and let $T \rightarrow \infty$ at the end.} Then, as depicted in figure (\ref{fig1})$,\mathcal{\bar{M}}=\mathcal{M}-\mathcal{B}\times\left[-T,T\right]$
denotes the spacetime region between the event horizon of the black
hole and the boundary of the black hole at infinity. Here $\mathcal{M}$ is the
total spacetime and $\mathcal{B}$ denotes the three dimensional ball
of which the boundary is the two dimensional cross section of the
event horizon. Also, $\partial\mathcal{\bar{M}}$ denotes the disconnected
boundary of that region: one at spatial infinity $\partial\mathcal{M}$,
the other on the event horizon $S^{2}\times\left[-T,T\right]$.
{{Under these considerations the hypersurface for asymptotically flat spacetimes is given as $\Sigma= \mathbb{R}^3-  \mathcal{B}$, with two a disconnected boundary composed of an $S^2$ as the cross section of the event horizon and another  $S^2$ at spatial infinity.}

The expression (\ref{integralequation}) is an identity for all Ricci flat metrics in four dimensions.
One important point to note is the following, for black holes the
Kretschmann scalar is divergent somewhere inside the event horizon
(defining the real singularity of the black hole); and its integral
over the totality of the spacetime is also clearly divergent, but
here we restrict the integration domain to the {{spatial}} region outside the
black hole for which the integral is finite.  Let us understand the content of the identity, (\ref{integralequation}),
in the case of the Kerr metric.

\subsection*{Application to the Kerr black hole}

The coordinates in which the metric is written does not change our
construction as we need the Kretschmann scalar, but since we also need
a Killing vector field, it is best to take coordinates in such a way
that one of them is a Killing coordinate. To this end,
one can take the Ricci flat Kerr metric in the Kerr-Schild form \cite{Schild} 
\begin{equation}
ds^{2}=d\bar{s}^{2}+\frac{2mr}{\rho^{2}}\left(k_{\mu}dx^{\mu}\right)^{2},
\label{kerrmet}
\end{equation}
where $\rho^{2}:=r^{2}+a^{2}\cos^{2}\theta$ and with the seed metric
given as 
\begin{eqnarray}
d\bar{s}^{2} & = & -dt^{2}+\frac{\rho^{2}dr^{2}}{\left(r^{2}+a^{2}\right)}+\rho^{2}d\theta^{2}+\left(r^{2}+a^{2}\right)\sin^{2}\theta d\phi^{2},
\end{eqnarray}
The vector $k_{\mu}$, {{which is null with respect to both the seed and the full metric,}} is given as 
\[
k_{\mu}dx^{\mu}=dt+\frac{\rho^{2}dr}{\left(r^{2}+a^{2}\right)}-a\sin^{2}\theta d\phi.
\]
{{For more details on the Kerr metric, see \cite{Visser}.}}
The outer event horizon is located at $r_{H}=m+\sqrt{m^{2}-a^{2}}$;
and the surface gravity, $\kappa$, at the event horizon can be easily
computed from the usual definition via the formula (\ref{surfacegravity}),
with the Killing vector field 
{
{\begin{equation}\label{zeta}
\zeta=\partial_{t}+\varOmega_{H}\partial_{\phi},
\end{equation}
which is the horizon-generating null Killing vector field.} 
Here $\varOmega_{H}$
is the angular velocity of the event horizon given as 
\begin{equation}
\varOmega_{H}=\frac{a}{r_{H}^{2}+a^{2}},
\end{equation}
{which makes $\zeta^{\mu} \zeta_\mu =0$ on the event horizon}

{{As mentioned in the paragraph below (\ref{surfacegravity}), $\zeta^{\mu}$ has a scaling ambiguity: the choice (\ref{zeta}) with the factor 1 in front of the timelike Killing vector removes this ambiguity which is the common practice that is consistent with the laws  of black hole mechanics.}} So, using (\ref{zeta}) in (\ref{surfacegravity}) one arrives at the known result \cite{Wald} for the surface gravity of the Kerr black hole
\begin{equation}
\kappa=\frac{r_{H}^{2}-a^{2}}{2r_{H}(r_{H}^{2}+a^{2})},\label{eq:surfacegravity}
\end{equation}
and the Hawking temperature follows from (\ref{Hawkingtemperature}).

Let us now show how our integral formula (\ref{integralequation}) reproduces
this result in a completely different manner. One can compute either
the left-hand side or the right-hand side of the identity (\ref{integralequation}), as they
are equal, the result of course does not matter. Defining the right-hand
side of (\ref{integralequation}) as
\begin{equation}
\mathcal{E}\left[\xi\right]:=-\frac{1}{4}\intop_{{\Sigma}}d^{3}y\sqrt{\gamma}n_\mu \xi^{\mu}R_{\rho\alpha\beta\sigma}R^{\rho\alpha\beta\sigma},
\label{exp}
\end{equation}
The Kretschmann scalar $K \equiv R_{\rho\alpha\beta\sigma}R^{\rho\alpha\beta\sigma}$ for the metric (\ref{kerrmet}) can be computed to be
\begin{eqnarray}
K = -96 m^2 \frac{A}{B}
\label{kresh}
\end{eqnarray}
where
\begin{eqnarray}
A= a^6 \cos (6 \theta )&+&10 a^6-180 a^4 r^2+240 a^2 r^4 +6 a^4 \left(a^2-10 r^2\right) \cos (4 \theta )\nonumber \\
&+&15 a^2 \left(a^4-16 a^2 r^2+16 r^4\right) \cos (2 \theta )-32 r^6,
\end{eqnarray}
and
\begin{equation}
B= \left(a^2 \cos (2 \theta )+a^2+2 r^2\right)^6.
\end{equation} 
{The induced metric $\gamma$ in the hypersurface $\Sigma$ follows from (\ref{kerrmet}) by setting $t=$constant, that is 
\begin{eqnarray}
d\bar{s}_\gamma^{2}  =\frac{\rho^{2}dr^{2}}{\left(r^{2}+a^{2}\right)}+\rho^{2}d\theta^{2}
 +\left(r^{2}+a^{2}\right)\sin^{2}\theta d\phi^{2},
\end{eqnarray}
and the vector $k_{\mu}$ reduces to
\[
k_{\mu}dx^{\mu}=\frac{\rho^{2}dr}{\left(r^{2}+a^{2}\right)}-a\sin^{2}\theta d\phi.
\]
}
{Taking the time-like Killing vector $\xi = (1,0,0,0)$ and computing the time-like unit normal 
$n_\mu$ to the hypersurface $\Sigma$ as
\begin{equation}
n_\mu = -\left(\frac{1}{\sqrt{1+\frac{4 m r \left(a^2+r^2\right)}{\left(a^2+r (r-2 m)\right) \left(a^2 \cos (2 \theta )+a^2+2 r^2\right)}}},0,0,0 \right)
\end{equation}
Then plugging all these into (\ref{exp}) and carrying out the volume integral over the ranges $r \in [r_H,\infty]$  and  $\theta \in [0,\pi]$,  $\phi \in [0,2\pi]$  that cover $\Sigma$, yields} 
\begin{equation}
\mathcal{E}\left[\partial_{t}\right]=-\frac{16\pi r_{H}m^{2}(r_{H}^{2}-a^{2})}{(r_{H}^{2}+a^{2})^{3}}.
\end{equation}
{This expression has the correct behavior for the surface gravity, for example it vanishes exactly for the extremal Kerr metric for which  $a=m$ and $r_H =m$. For the subextremal Kerr metric, one must introduce a constant coefficient which is akin to the scaling ambiguity in (\ref{surfacegravity}). In fact, this is evident from the choice of the Killing vector above: we chose $\xi = (1,0,0,0)$, but any other choice $\tilde{\xi} = (c,0,0,0)$, with $c$ a constant, would scale $\kappa$. }
So one has the surface gravity of the Kerr metric
\begin{equation}
\kappa=-\frac{1}{32\pi}\left(\frac{a}{mr_{H}\Omega_{H}}\right)^{2}\mathcal{E}\left[\partial_{t}\right],\label{kappa}
\end{equation}
which is equivalent to (\ref{eq:surfacegravity}). For the Schwarzschild
black hole, $a=0$ and one finds the correct limit $\kappa=\frac{1}{4m}$.

 Defining
the dimensionless rotation parameter of the black hole as $\alpha:=a/m$
and the tangential speed of the horizon as $v_{H}=r_{H}\Omega_{H}$,
the prefactor in (\ref{kappa}) reads as the dimensionless ratio $(\alpha/v_{H})^{2}$. so
for the Kerr black hole, the Hawking temperature simply reads as 

\noindent\fbox{\begin{minipage}[t]{1\columnwidth - 2\fboxsep - 2\fboxrule}%
\begin{equation}
T_{H}=\left(\frac{a}{16\pi mr_{H}\Omega_{H}}\right)^{2}\intop_{{\Sigma}}d^{3}y\sqrt{\gamma}n_\mu \xi^{\mu}R_{\rho\alpha\beta\sigma}R^{\rho\alpha\beta\sigma},
\end{equation}
\end{minipage}}\\
\\
where, again, the integral is outside the black hole region. Equivalently,
from (\ref{integralequation}) one can calculate the same integral
on the surface of the event horizon.
\begin{figure}
~~\includegraphics[scale=0.7]{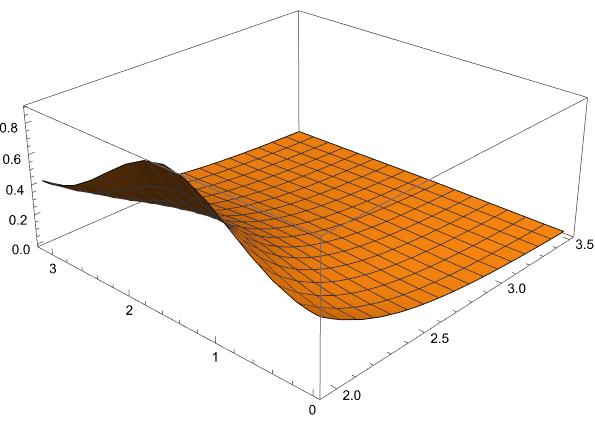}
\caption{Plot of the Kretschmann scalar $K$ (\ref{kresh}) for the Kerr black hole with $a=1/3$, $m=1$; and for the interval $ r\in [r_H=1.94,3.5]$, $\theta \in [0,\pi]$ . The figure is depicted to show how the total  Kretschmann scalar of the rotating black hole over the spatial region outside the black hole region can give a finite result.
}
\label{fig2}
\end{figure}

{In Figure (\ref{fig2}), we have plotted the  Kretschmann scalar $K$ (\ref{kresh}) for the Kerr black hole. Even though this curvature scalar diverges on a ring inside the event horizon, diagnosing the true singularity of spacetime, its finite outside the event horizon. Moreover, as is the premise of this work, as we have shown, its spatial volume integral is also finite and is related to the surface gravity. }

{Let us give a non-vacuum example: the extremal charged Reissner-Nordstr\"om metric which is a solution to Einstein-Maxwell theory. The metric is the of the following form
\begin{equation}
ds^2 = -f(r) dt^2 + \frac{ dr^2}{f(r)} + r^2( d\theta^2 + \sin^2\theta d\phi^2)
\label{RN}
\end{equation}
where we keep $f(r)$ arbitrary for now but will specify in a moment. The metric is not Ricci-flat generically, so we must now apply the full identity (\ref{mainidentity}). We assume $\Lambda=0$ and in four dimensions, we we already noted, the Gauss-Bonnet 2-tensor vanishes, but the Gauss-Bonnet scalar does not.  For (\ref{RN}), we have
\begin{equation}
\chi_{GB}=\frac{4 \left((f-1) f''+\left(f'\right)^2\right)}{r^2}
\end{equation}
for $f(r) = 1- 2m/r + q^2/r^2$, for the time-like vector $\xi = (1,0,0,0)$, the relevant expression  
\begin{equation}
\mathcal{E}\left[\xi\right]:=-\frac{1}{4}\intop_{{\Sigma}}d^{3}y\sqrt{\gamma}n_\mu \xi^{\mu}\chi_{GB},
\label{exp2}
 \end{equation}
yields
\begin{equation}
\mathcal{E}\left[\xi\right]=\frac{8 \pi}{r_H^5} \left(2 r_H^2 m^2 - 3 r_H m q^2 + q^4\right),
\end{equation}
which vanished identically for the extremal case of $q=m$, that is $r_H = m$. This yield $\kappa =0$ and so $T_H=0$ as expected. For the non-extremal case, the constant in front of the Killing vector must be taken not 1.}

Finally, let us also note a rather interesting connection (which needs to be better studied) of the above construction with our earlier work \cite{Emel_PRD_uzun}, \cite{Altas}. The {\it linearized } version
of the main identity (\ref{mainidentity}) yields the conserved charges
(mass and angular momentum) written in terms of not the derivatives
of the metric deviations but in terms of the explicitly diffeomorphism
invariant linearized Riemann tensor for asymptotically anti-de Sitter
spacetimes \cite{Emel_PRD_uzun}, \cite{Altas} as
\begin{equation}
Q\left(\bar{\xi}\right)=k\int_{\partial\bar{\Sigma}}d^{n-2}x\,\sqrt{\bar{\gamma}}\,\bar{\epsilon}_{\mu\nu}\left(R^{\nu\mu}\thinspace_{\beta\sigma}\right)^{\left(1\right)}\bar{\text{\ensuremath{{\cal {F}}}}}^{\beta\sigma},\label{newcharge}
\end{equation}
with the constant coefficient $k=(n-1)(n-2)/[8(n-3)\Lambda G\Omega_{n-2}]$
and the barred quantities refer to the AdS background; {and $2\bar{\text{\ensuremath{{\cal {F}}}}}_{\beta\sigma}= \bar{\nabla}_\beta \bar{\xi}_\sigma -\bar{\nabla}_\sigma \bar{\xi}_\beta$.}

\section{Conclusions}
We have given two new expressions for surface gravity and the associated temperature for black holes. Broadly speaking both surface gravity and the temperature of a black hole is directly related to the  total (integrated) quadratic curvature invariant outside the black hole region. This is a rather important result which directly links the surface gravity to a curvature invariant and the integration clearly shows the non-locality of the surface gravity concept.  

Our construction was based on a divergence-free rank four tensor and an antisymmetric rank
two tensor built from the covariant derivative of any Killing vector
field: we found an identity (\ref{mainidentity}) valid for all
spacetimes of generic $n>3$ dimensions. A curious fact is that the
Gauss-Bonnet tensor $\mathcal{H}_{\mu\nu}$ (which comes from the
variation $\delta_{g}\intop d^{n}x\sqrt{-g}\chi_{GB}$) and the Gauss-Bonnet
scalar, $\chi_{GB}$, appears on the right-hand side of the expression.
When this expression is integrated in a region of spacetime for which
the integrals are finite, one obtains an identity, which in four dimensional
Ricci flat metrics yield (\ref{integralequation}). Remarkably the
left-hand side or the right-hand side of this identity is related
to the Hawking temperature (and the surface gravity) of the black
hole. We realized this upon the computation for extremal black holes;
the integrals vanish identically and for Schwarzschild black holes
they yield the surface gravity $\kappa=1/4m$ which are the known results.  Therefore, besides the usual way
of defining the surface gravity via the null geodesic generator as
in (\ref{surfacegravity}), our construction gives two novel definitions
one of which includes the integral of the Kretschmann scalar in the
part of the spacetime outside the black hole region, and the other
one being an integral of the Riemann tensor (in the Ricci flat case
it is actually the Weyl tensor) contracted with the covariant derivative
of any Killing vector field.

\end{document}